\documentclass[fleqn,usenatbib]{mnras}

\usepackage{mathptmx}
\usepackage{txfonts}

\usepackage[T1]{fontenc}
\usepackage{ae,aecompl}

\usepackage{graphicx}	

\usepackage{float}

\newcommand{\beq}{\begin{equation}}
\newcommand{\eeq}{\end{equation}}
\newcommand{\beqa}{\begin{eqnarray}}
\newcommand{\eeqa}{\end{eqnarray}}

\newcommand{\NH}{N_{\rm{H}}}

\newcommand{\Fsoft}{F_{0.25-2,{\rm unabs}}}
\newcommand{\Ftot}{F_{0.25-8,{\rm unabs}}}

\newcommand{\Tbb}{T_{\rm{bb}}}
\newcommand{\Tin}{T_{\rm{in}}}

\newcommand{\Lx}{L_{\rm{X}}}

\newcommand{\Lunabs}{L_{{\rm X,unabs}}}
\newcommand{\Lunabsj}{L_{{\rm X,unabs},j}}
\newcommand{\Lobs}{L_{{\rm X,obs}}}

\newcommand{\LSunabs}{L_{{\rm SX,unabs}}}

\newcommand{\LSobs}{L_{{\rm SX,obs}}}

\newcommand{\Lunabsi}{L_{i,{\rm unabs}}}
\newcommand{\Lunabsij}{L_{i,{\rm unabs},j}}
\newcommand{\Lobsi}{L_{i,{\rm obs}}}
\newcommand{\Lobsij}{L_{i,{\rm obs},j}}

\newcommand{\Sunabsi}{S_{i,{\rm unabs}}}
\newcommand{\Sobsi}{S_{i,{\rm obs}}}

\newcommand{\FLMXB}{f_{{\rm LMXB}}}
\newcommand{\Cvar}{C_{{\rm var}}}

\title[Collective spectrum of luminous HMXBs]{The intrinsic collective
  X-ray spectrum of luminous high-mass X-ray binaries} 
 
\author[S. Sazonov and
  I. Khabibullin]{S. Sazonov$^{1,2}$\thanks{E-mail:
    sazonov@iki.rssi.ru} and I. Khabibullin$^{3,1}$\\ 
$^{1}$Space Research Institute, Russian Academy of Sciences,
  Profsoyuznaya 84/32, 117997 Moscow, Russia\\  
$^{2}$Moscow Institute of Physics and Technology, Institutsky per. 9,
  141700 Dolgoprudny, Russia \\ 
$^{3}$Max Planck Institute for Astrophysics,
  Karl-Schwarzschild-Strasse 1, D-85741 Garching, Germany 
}

\begin{document}
\label{firstpage}
\pagerange{\pageref{firstpage}--\pageref{lastpage}}
\maketitle

\begin{abstract}
  Using a sample of two hundred luminous ($\Lunabs>10^{38}$~erg~s$^{-1}$,
  where $\Lunabs$ is the unabsorbed 0.25--8~keV luminosity) high-mass
  X-ray binary (HMXB) candidates found with {\sl Chandra} in 27
  nearby galaxies, we have constructed the collective X-ray
    spectrum of HMXBs in the local Universe 
  per unit star formation rate, corrected for observational biases
  associated with intrinsic diversity of HMXB spectra and X-ray
  absorption in the interstellar medium. This spectrum is well
  fit by a power law with a photon index 
  $\Gamma=2.1\pm 0.1$ and is dominated by ultraluminous X-ray sources
  with $\Lunabs>10^{39}$~erg~s$^{-1}$. Hard sources (those with the
  0.25--2~keV to 0.25--8~keV flux ratio of $<0.6$) dominate above
  $\sim 2$~keV, while soft and supersoft sources (with the flux ratios
  of 0.6--0.95 and $>0.95$, respectively) at lower
  energies. The derived spectrum probably represents the
  angle-integrated X-ray emission of the near- and super-critically
  accreting stellar mass black holes and neutron stars in the local
  Universe. It provides an important constraint on supercritical
    accretion models and can be used as a reference spectrum for
  calculations of the X-ray preheating of the Universe by the first
  generations of X-ray binaries.
\end{abstract}


\section{Introduction}
\label{s:intro}

The X-ray emission of actively star forming galaxies is dominated by
the collective signal of high-mass X-ray binaries (HMXBs),
complemented by diffuse soft X-ray emission from hot interstellar gas
(e.g. \citealt{lehetal10}). Although the HMXB population of the Milky 
Way has been thoroughly studied (see \citealt{waletal15} for a recent 
review), the most luminous X-ray binaries are so rare that they can be
found and counted only in other (nearby) galaxies. Such studies have
revealed that the HMXB X-ray luminosity function (XLF) has a
power-law shape: $dN/d\Lx\propto\Lx^{-1.6}$ from $\Lx\lesssim 10^{36}$
to $\sim 10^{40}$~erg~s$^{-1}$ \citep{minetal12a}, with some indication
of steepening at $\Lx\gtrsim 10^{40}$~erg~s$^{-1}$. The bulk of the
emission from point X-ray sources in actively star forming galaxies is
thus produced by ultraluminous X-ray sources (ULXs) with $\Lx\gtrsim
10^{39}$~erg~s$^{-1}$, the majority of which appear to be
supercritically accreting stellar-mass black holes
(e.g. \citealt{pouetal07,fensor11,robetal16}) and neutron stars
\citep{bacetal14,musetal15,fueetal16,isretal17}. 

Luminous HMXBs, including ULXs and so-called ultraluminous supersoft
sources (ULSs), exhibit a large variety of X-ray spectral shapes,
probably due to differences in the accretion rate and orientation of
the thick accretion disc with respect to the observer
(e.g. \citealt{glaetal09,sutetal13,urqsor16}) as well as in the nature
of the accretor (a neutron star vs. a black hole,
\citealt{pinetal17}). In addition, the observed X-ray spectra and
detection rates of such objects can be significantly affected by
photoabsorption in the interstellar medium (ISM) of the host galaxies
and of the Milky Way. Taking all this into account, we have 
recently constructed \citep{sazkha17a} the {\sl intrinsic} HMXB XLF in
its bright end, $10^{38}<\Lunabs\lesssim 10^{40.5}$~erg~s$^{-1}$ (where
$\Lunabs$ is the absorption corrected source luminosity in the
0.25--8~keV energy band), per unit star formation rate (SFR). It can
be described by a power law, $dN/d\log\Lunabs\approx
2.0(\Lunabs/10^{39}\,{\rm
  erg~s}^{-1})^{-0.6}$~$(M_\odot$~yr$^{-1})^{-1}$, which has the same
slope as the {\sl observed} HMXB XLF of \citet{minetal12a} but a factor
of $\sim 2.3$ higher normalization. We further showed that about two 
thirds of the total X-ray (0.25--8~keV) emission of HMXBs is released
in the soft band (0.25--2~keV), $\sim 5\times
10^{39}$~erg~s$^{-1}$~$(M_\odot$~yr$^{-1})^{-1}$, with roughly equal
contributions from 'hard', 'soft' and 'supersoft' sources, defined 
according to their intrinsic soft/total X-ray flux ratio: 
\begin{eqnarray}
{\rm Hard:}& \Fsoft/\Ftot\le 0.6,\nonumber\\
{\rm Soft:}& 0.6<\Fsoft/\Ftot\le 0.95,\nonumber\\
{\rm Supersoft:}& \Fsoft/\Ftot>0.95.
\label{eq:types}
\end{eqnarray}

This detailed information about the intrinsic XLF provides
interesting constraints on the population properties of HMXBs and the 
physics of near- and super-critical accretion. It may also be
interesting in the context of studying the 'cosmic dawn', since HMXBs
belonging to the first generations of stars and their remnants might
have radiatively preheated the Universe before it was reionized by
UV radiation from stars and quasars (e.g. \citealt{miretal11}). In 
our other recent paper \citep{sazkha17b}, we demonstrated (see also
\citealt{madfra16}) that HMXBs
could significantly heat the Universe at $z\sim 10$ if the specific
(i.e. per unit SFR) X-ray emissivity of such systems was higher by an
order of magnitude than at the present epoch and the soft X-rays
produced by HMXBs could escape from their host galaxies without strong
attenuation in their ISM. Whether or not these conditions were
fulfilled is an open question. 

In \citet{sazkha17b} we used the measured ratio of the HMXB luminosity
functions in the 0.25--2 and 0.25--8~keV bands to 
estimate the effective photon index of the average intrinsic X-ray
spectrum of luminous HMXBs: $\Gamma\sim 2.1$. In the present study, we
take advantage of the same sample of sources to construct the
  intrinsic (i.e. corrected for observational biases), SFR-normalized
  energy spectrum of the integrated emission of luminous HMXBs in the
  local Universe, hereafter referred to as the {\sl intrinsic collective
    spectrum of HMXBs}, and evaluate the contributions of hard, soft and
supersoft sources to it. This spectrum may find application, in
particular, in simulations of the preheating of the early Universe by
X-ray binaries.

\section{Sample}
\label{s:sample}

We make use of the 'clean sample' of X-ray sources detected by the
{\sl Chandra} X-ray Observatory \citep{wanetal16}, presumably located
in 27 nearby ($D<15$~Mpc) galaxies (mostly spirals), from
\citet{sazkha17a}. This sample had been compiled based on the following  
criteria: i) the source must be located on the sky within the 25 mag
arcsec$^{-2}$ isophote of the galaxy, i.e. at radius $R<R_{25}$ in the
plane of the galaxy, but outside of its central $0.05 R_{25}$ region,
ii) there must be at least 100 photon counts from the source in some
{\sl Chandra} observation and iii) the unabsorbed
0.25--8~keV luminosity of the source must exceed
$10^{38}$~erg~s$^{-1}$. Having additionally filtered out 16 known or
suspected background active galactic nuclei (AGN) and 3 foreground
Galactic stars, we had selected 200 HMXB candidates with
luminosities ranging from $\Lunabs=10^{38}$ to $\sim 3\times
10^{40}$~erg~s$^{-1}$, and analyzed their {\sl Chandra} spectra.  

\section{Analysis}
\label{s:analysis}

In \citet{sazkha17a}, we described the measured spectrum of each
source by one of the following models: i) absorbed power law, ii)
absorbed blackbody emission and iii) absorbed mutlicolour blackbody
disc emission. These best-fitting models were then used to determine
the sources' intrinsic and observed luminosities in the 0.25--8~keV
and 0.25--2~keV energy bands ($\Lunabs$, $\Lobs$, $\LSunabs$ and 
$\LSobs$, respectively). We now use the same spectral
fits to determine the intrinsic and observed luminosities 
($\Lunabsi$ and $\Lobsi$, respectively) in 5 narrow subbands:
0.25--0.5, 0.5--1, 1--2, 2--4, and 4--8~keV (hereafter referred to as
bands $i=1$ to 5).

Using high-quality maps of atomic (HI) and molecular (H$_2$)
  gas in the sampled galaxies, we demonstrated in \citet{sazkha17a}
  that the line-of-sight absorption column densities, $\NH$, inferred
  from the spectra of the studied X-ray sources (typically a few
  $10^{21}$~cm$^{-2}$) can be attributed to the ISM of
  their host galaxies. We further took advantage of the HI,
  H$_2$ and SFR maps of the sampled galaxies to evaluate observational
  biases associated with the detection of X-ray sources by {\sl
    Chandra}, which arise due to intrinsic diversity of HMXB
  spectra and X-ray absorption in the ISM. In a nutshell (see
  \citealt{sazkha17a} for a detailed discussion), in the absence of 
  intervening absorbing gas, a soft source with a given intrinsic
  luminosity $\Lunabs$ would produce more photon counts on the
  detector than a hard source with the same luminosity and
  location. On the other hand, observed X-ray fluxes of soft sources are more
  strongly affected by photoabsorption in the ISM, so that such
  sources may become hidden from {\sl Chandra} if located on the 
  farther side of their host galaxy. As a result, only some fraction of the
  total SFR ($\sim 32$~$M_\odot$~yr$^{-1}$) in the 27 sampled galaxies
  is effectively probed by {\sl Chandra} in X-rays, and this
  fraction depends on the source intrinsic luminosity and spectral
  type, $T$:  ${\rm SFR~}(\Lunabs,T)$ \citep{sazkha17a}.  

In \citet{sazkha17a}, we used the ${\rm SFR~}(\Lunabs,T)$
dependence to derive the HMXB XLF per unit SFR. We can now apply a
similar procedure to constuct the intrinsic collective spectrum of
HMXBs in the local Universe by summation over the sources in the sample: 
\beq
\Sunabsi=\Cvar\sum_j[1-\FLMXB(\Lunabsj)]\frac{\Lunabsij}{{\rm
    SFR~}(\Lunabsj,T_j)}. 
\label{eq:avspec_intr}
\eeq
Here, $\Sunabsi$ is the HMXB emissivity [measured in units of
  erg~s$^{-1}$~($M_\odot$~yr$^{-1}$)$^{-1}$] in energy band $i$ (from
1 to 5), $\Lunabsij$ is the intrinsic luminosity in band $i$ of the
$j$th source, $\Lunabsj$ is its intrinsic 0.25--8~keV luminosity and $T_j$
is its spectral type.

By means of equation~(\ref{eq:avspec_intr}), we perform a weighted
stacking of the source spectra, such that each {\sl Chandra} source
provides a contribution equal to its luminosity in a given energy band
divided by the corresponding X-ray-probed SFR. This procedure is
analogous to the $1/V_{\rm max}$ weighting method frequently used in
astronomy, with ${\rm SFR}$ playing the role of a generalized $V_{\rm
  max}$. Similar procedures have been used before e.g. for estimating
the space density of stellar objects in the Galaxy taking into 
account their inhomogeneous spatial distribution
(e.g. \citealt{tinetal93,sazetal06}). Our current treatment also
closely follows the calculation of the collective X-ray spectrum of
AGN in the local Universe by \citet{sazetal08}.

We have added two additional factors in
equation~(\ref{eq:avspec_intr}). The factor $1-\FLMXB(\Lunabsj)$ takes
into account that our sample of HMXB candidates is expected to be
contaminated by low-mass X-ray binaries (LMXBs). Their relative
contribution as a function of luminosity was estimated in
\citet{sazkha17a} based on the LMXB XLF \citep{gilfanov04} and can be
approximated as
\beqa
\FLMXB =
\left\{
\begin{array}{ll}
0.49, & 10^{38}\le\Lunabs<10^{38.5}~{\rm erg~s}^{-1},\\
0.41, & 10^{38.5}\le\Lunabs<10^{39}~{\rm erg~s}^{-1},\\
0.06, & 10^{39}\le\Lunabs<10^{39.5}~{\rm erg~s}^{-1},\\
0, & \Lunabs\ge10^{39.5}~{\rm erg~s}^{-1}.
\end{array}
\right.
\label{eq:lmxb}
\eeqa
The LMXB contribution is thus substantial below $10^{39}$~erg~s$^{-1}$
but negligible at higher luminosities. Therefore, since most of the
emission from HMXBs is produced by ULXs (with
$\Lunabs>10^{39}$~erg~s$^{-1}$), the significant uncertainty in our
knowledge of the LMXB XLF and hence their contribution to our sample
does not translate into a significant uncertainty in the resulting
collective spectrum of HMXBs.

The additional coefficient $\Cvar=1/1.2\approx 0.83$ in
equation~(\ref{eq:avspec_intr}) takes into account the 'variability
bias' evaluated in \citet{sazkha17a}. It results from our using 
particular {\sl Chandra} observations for estimating the luminosities
of the sources (namely those with at least 100 counts from the source)
while the same sources would be weaker by $\sim 20$\% on average due
to intrinsic variability if their luminosities were measured randomly
in time. 

There are two types of uncertainties associated with the 
  collective spectrum $\Sunabsi$. One is due to uncertainties,
$\delta\Lunabsij$, in estimation of the unabsorbed luminosities of the
inidividual sources from X-ray spectral analysis: 
\beq
\delta_{i,1}=\Cvar\sqrt{\sum_j\left([1-\FLMXB(\Lunabsj)]\frac{\delta\Lunabsij}{{\rm
      SFR~}(\Lunabsj,T_j)}\right)^2}.
\label{eq:avspec_unc1}
\eeq
Another arises from the finite size of our source sample: 
\beq
\delta_{i.2}=\Cvar\sqrt{\sum_j\left([1-\FLMXB(\Lunabsj)]\frac{\Lunabsij}{{\rm
      SFR~}(\Lunabsj,T_j)}\right)^2}.
\label{eq:avspec_unc2}
\eeq
This uncertainty stems from the fact that in the standard $\sum
1/V_{{\rm max},j}$ estimation of space densities, the uncertainty of
each object's contribution is assumed to follow Poisson statistics
\citep{feletal76} so that the variance of the density estimate is
$\sum 1/V^2_{{\rm max},j}$ (e.g. \citealt{tinetal93}). In our case,
the contributions of individual sources to the variance of $\Sunabsi$
are similarly independent of each other but must be multiplied by the
square of the corresponding coefficients in
equation~(\ref{eq:avspec_intr}). 

The total uncertainty can be estimated as a combination of these
uncertainties: 
\beq
\delta_i=\sqrt{\delta_{i,1}^2+\delta_{i,2}^2}.
\label{eq:avspec_unc}
\eeq

We can use a slightly modified stacking procedure to also compute the 
{\sl observed} collective spectrum of HMXBs in the
local Universe:
\beq
\Sobsi=\Cvar\sum_j[1-\FLMXB(\Lunabsj)]\frac{\Lobsij}{{\rm
    SFR~}(\Lunabsj,T_j)}.
\label{eq:avspec_obs}
\eeq
This spectrum represents the integrated X-ray emission of HMXBs as
seen by the Earth's observer, i.e. uncorrected for line-of-sight
absorption. In this case, the uncertainties of the first type are
negligible due to the small $\delta\Lobsij$ errors and those of the
second type can be computed using equation~(\ref{eq:avspec_unc2}) by
substituting $\Lobsij$ for $\Lunabsij$. 

\section{Results}
\label{s:results}

\subsection{Intrinsic spectrum}
\label{s:avspec_intr}

\begin{figure}
\centering
\includegraphics[width=\columnwidth,viewport=30 200 560
  710]{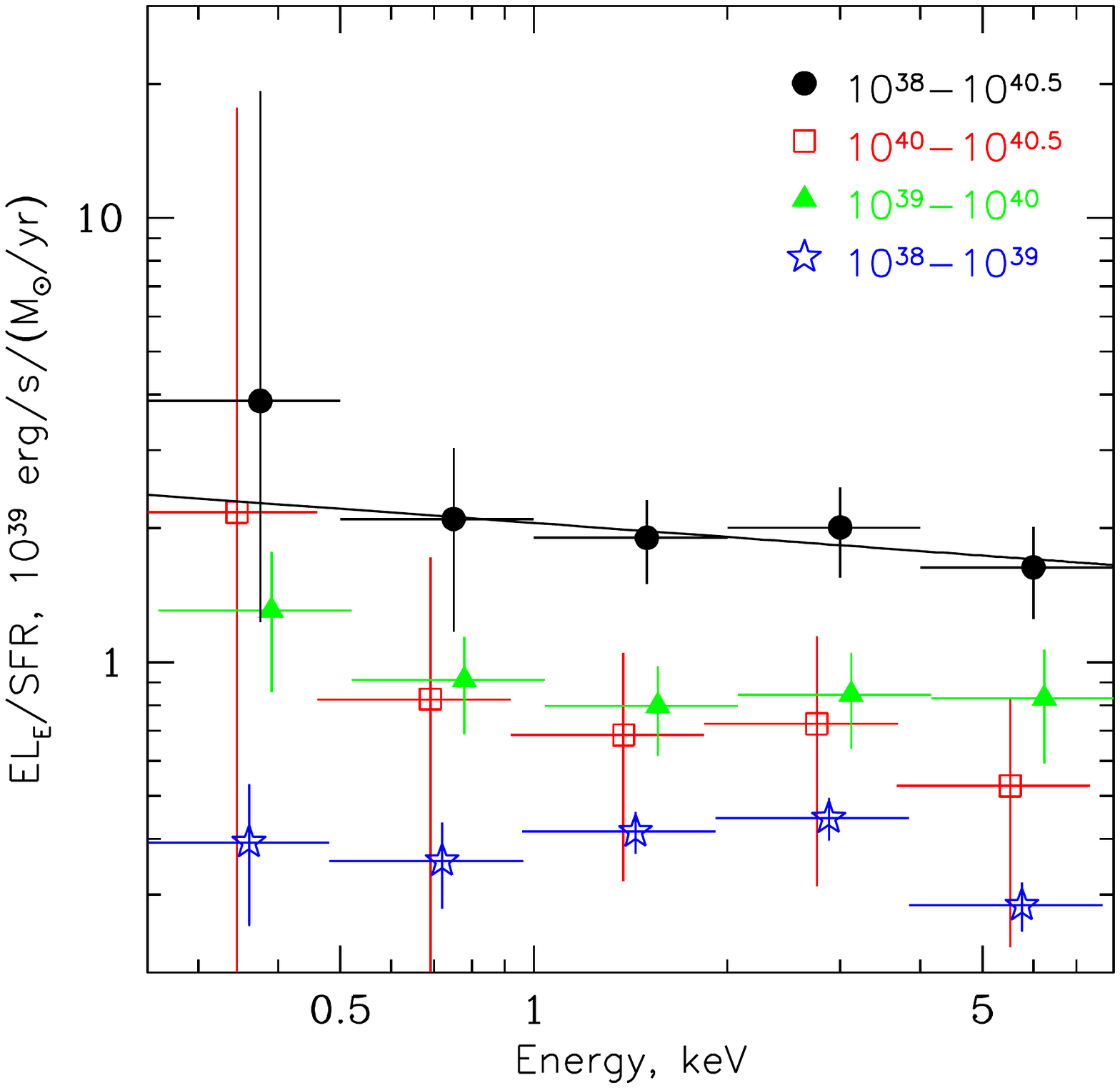} 
\caption{Intrinsic collective spectrum of HMXBs in the local
  Universe (black circles) and the contributions  
  (these points are slightly shifted along the horizontal axis for
  better visibility) of sources in three luminosity ranges:
  $\Lunabs=10^{40}$--$10^{40.5}$~erg~s$^{-1}$ (red squares),
  $10^{39}$--$10^{40}$~erg~s$^{-1}$ (green triangles) and
  $10^{38}$--$10^{39}$~erg~s$^{-1}$ (blue stars). The solid line is
  the best-fitting power law for the total spectrum
  [eq.~(\ref{eq:fit_l38_40.5})].
}  
\label{fig:avspec_lumranges}
\end{figure}

Figure~\ref{fig:avspec_lumranges} shows the {\sl intrinsic} 
  collective spectrum of HMXBs with $10^{38}<\Lunabs\lesssim
10^{40.5}$~erg~s$^{-1}$ (the luminosity range spanned by our sample
of sources), obtained using
equation~(\ref{eq:avspec_intr}). It can be well fitted ($\chi^2=0.21$
for 3 degrees of freedom) by a power law: 
\beq
\frac{EL_E}{\rm SFR}=(2.1\pm 0.4)\times 10^{39} \left(\frac{E}{{\rm
    keV}}\right)^{-0.11\pm 0.18}\,{\rm erg~s}^{-1}\,(M_\odot~{\rm yr})^{-1}.
\label{eq:fit_l38_40.5}
\eeq
The quoted uncertainty for the spectral slope may be slightly 
overestimated because we regard the uncertainties $\delta_i$ of 
the individual spectral points as independent, although they may be
somewhat correlated due to the contribution $\delta_{i,2}$
[eq.~(\ref{eq:avspec_unc2})] from the Poisson uncertainty in the 
number of sampled sources (this is only important for the three higher
energy channels, since the uncertainties in the 0.25--0.5~keV and
0.5--1~keV bands are dominated by the $\delta_{i,1}$ errors associated
with luminosity estimation for individual sources). 

As could be expected from the HMXB XLF \citep{sazkha17a},
the bulk of the emission from HMXBs is provided by
ultraluminous ($\Lunabs>10^{39}$~erg~s$^{-1}$) sources (see
Fig.~\ref{fig:avspec_lumranges}). Moreover, there is an indication
that a sizeable or even dominant contribution is provided by extremely
luminous sources with $\Lunabs>10^{40}$~erg~s$^{-1}$. These results
are unlikely to be strongly affected by LMXB contamination of our
sample of sources, which we have roughy taken into account through the
$1-\FLMXB$ factor in equation~(\ref{eq:avspec_intr}). Indeed, LMXBs
are only important at $\Lunabs<10^{39}$~erg~s$^{-1}$, but the overall
contribution of such relatively low-luminosity sources to the total
emission from HMXBs is small, while nearly all of our
$\Lunabs>10^{39}$~erg~s$^{-1}$ sources are expected to be HMXBs.

The large uncertainties of the collective spectrum of HMXBs in
the two softest 
bands are mainly associated with the presence of two very luminous
($\Lunabs\gtrsim 10^{40}$~erg~s$^{-1}$) supersoft [per our definition,
eq.~(\ref{eq:types})] sources in our sample. As discussed in
\citet{sazkha17a}, these sources have very soft spectra (which can be
described as blackbody radiation with $kT_{\rm bb}\sim
0.06$--0.07~keV) and their inferred intrinsic luminosities are some 3
orders of magnitude higher than their observed luminosities
(apparently due to the presence of significant amounts of cold ISM in
their direction) but very uncertain (by 1--2 orders of
magnitude). Moreover, there are in total only 7 sources (2 hard, 3
soft and 2 supersoft ones) with $\Lunabs>10^{40}$~erg~s$^{-1}$ in our
sample, so that the overall contribution of such luminous sources to
the total X-ray emission produced by HMXBs is not well
constrained by the present study.

\begin{figure}
\centering
\includegraphics[width=\columnwidth,viewport=30 200 560 710]{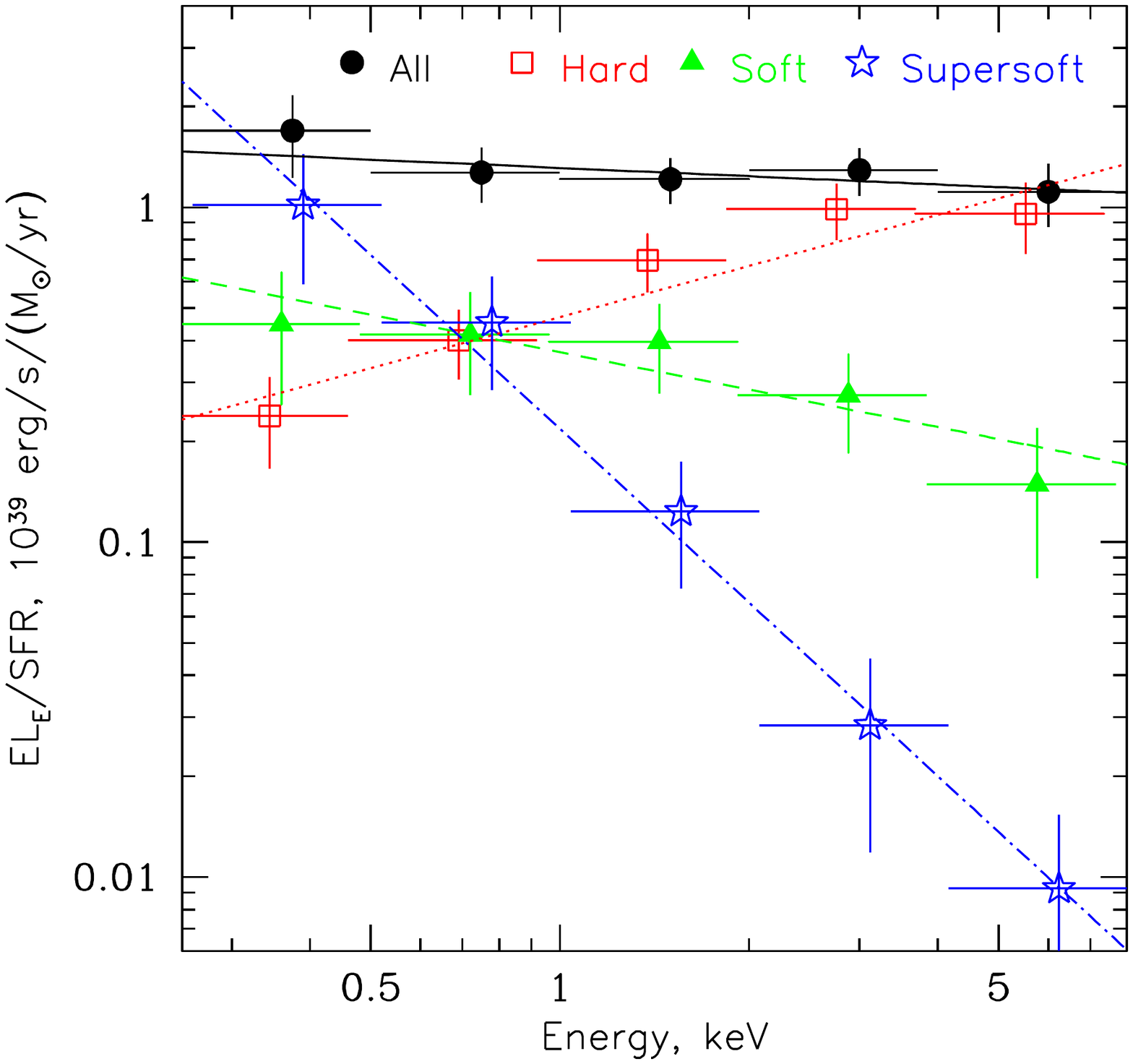} 
\caption{Intrinsic collective spectrum of HMXBs with
  $10^{38}<\Lunabs<10^{40}$~erg~s$^{-1}$ per unit SFR (black circles)
  and the contributions of hard, soft and supersoft sources
  [according to our definition, eq.~(\ref{eq:types})]: squares, triangles
  and stars, respectively. The black solid line shows the best-fitting
  power law for the total spectrum [eq.~(\ref{eq:fit_l38_40})]. The
  red dotted, green dashed and blue dash-dotted lines 
  show the corresponding fits for the hard, soft and supersoft 
  components [eqs.~(\ref{eq:fit_hard_l38_40}),
    (\ref{eq:fit_soft_l38_40}) and (\ref{eq:fit_supersoft_l38_40}),
    respectively]. 
} 
\label{fig:avspec_spectypes_fits}
\end{figure}

Given the large uncertainty associated with the contribution of the
most luminous sources ($\Lunabs>10^{40}$~erg~s$^{-1}$), we also
calculated the collective spectrum of HMXBs with
$10^{38}<\Lunabs<10^{40}$~erg~s$^{-1}$ (there are in total 193 such
objects in our sample), which is shown in
Fig.~\ref{fig:avspec_spectypes_fits}. This spectrum is tightly
constrained and can be well fitted ($\chi^2=0.71$ for 3 degrees of
freedom) by the following power law:
\beq
\frac{EL_E}{\rm SFR}=(1.31\pm 0.13)\times 10^{39}  \left(\frac{E}{{\rm
    keV}}\right)^{-0.08\pm 0.11}\,{\rm erg~s}^{-1} (M_\odot~{\rm yr})^{-1}.
\label{eq:fit_l38_40}
\eeq
Comparing this expression with equation~(\ref{eq:fit_l38_40.5}) we see
that the slope is unchanged, but the normalization has decreased 
by $\sim 20$--50\%, which reflects the substantial contribution of the
most luminous ($\Lunabs\gtrsim 10^{40}$~erg~s$^{-1}$) sources to the
total X-ray emission from HMXBs. The derived spectral slope
(photon index) $\Gamma\approx 2.1$ confirms the conclusion of our
previous work \citep{sazkha17a,sazkha17b} that about two thirds of the
total X-ray output of HXMBs emerges in the form of soft X-rays, at
energies below 2~keV.

\begin{figure}
\centering
\includegraphics[width=\columnwidth,viewport=30 200 560 710]{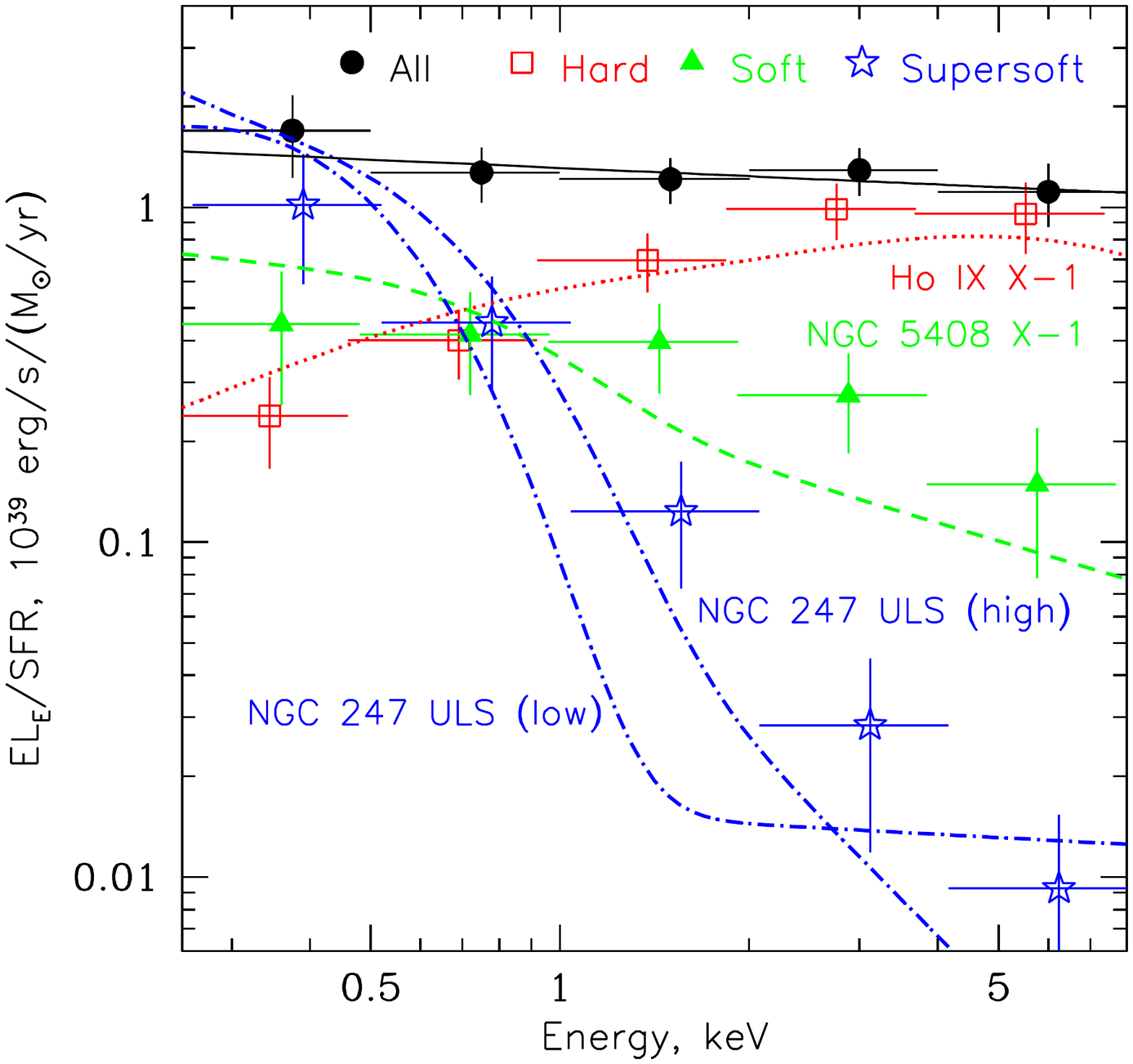} 
\caption{Same as Fig.~\ref{fig:avspec_spectypes_fits}, but the
  power-law fits for hard, soft and supersoft sources are replaced by
  examples of spectra (best-fitting unabsorbed models adopted from the
  literature, see main text, with ad hoc normalizations) of real
  individual sources belonging to tese catagories: red dotted line --
  ULX~Ho~IX~X-1, green dashed line -- ULX~NGC~5408~X-1, blue
    dash-dotted lines -- NGC~247~ULS in two different states (see
    references in the text). None of
  these sources belong to our sample.
} 
\label{fig:avspec_spectypes_examples}
\end{figure}

Although the shape of the collective spectrum of HMXBs is
consistent with a simple power law, this is merely the result of
summing over a great variety of individual spectra. In reality, as
shown in Fig.~\ref{fig:avspec_spectypes_fits}, hard, soft and
supersoft sources (again, per our definition) provide comparable
contributions to the total spectrum, and it is the soft and supersoft
sources that are largely responsible for its low-energy part. The
presented collective spectra of the hard, soft and supersoft
sources have been obtained using the same stacking procedure
[eq.~(\ref{eq:avspec_intr})] as for the total spectrum, applied to
117, 47 and 29 sources of these classes, respectively (excluding the 7
sources with $\Lunabs>10^{40}$~erg~s$^{-1}$). These spectra can be
well fitted ($\chi^2=2.79$, 1.06 and 0.63, respectively, for 3 degrees
of freedom) by the following power laws:
\beq
\left(\frac{EL_E}{\rm SFR}\right)_{{\rm hard}}=(0.47\pm 0.05)\times 10^{39} \left(\frac{E}{{\rm
    keV}}\right)^{0.51\pm 0.10}\,{\rm erg~s}^{-1} (M_\odot~{\rm yr})^{-1}.
\label{eq:fit_hard_l38_40}
\eeq
\beq
\left(\frac{EL_E}{\rm SFR}\right)_{{\rm soft}}=(0.37\pm 0.06)\times 10^{39} \left(\frac{E}{{\rm
    keV}}\right)^{-0.37\pm 0.16}\,{\rm erg~s}^{-1} (M_\odot~{\rm yr})^{-1}.
\label{eq:fit_soft_l38_40}
\eeq
\beq
\left(\frac{EL_E}{\rm SFR}\right)_{{\rm supersoft}}=(0.22\pm 0.09)\times 10^{39} \left(\frac{E}{{\rm
    keV}}\right)^{-1.72\pm 0.21}\,{\rm erg~s}^{-1} (M_\odot~{\rm yr})^{-1}.
\label{eq:fit_supersoft_l38_40}
\eeq

In reality our partition of HMXBs into three classes is ad hoc, and
each of these groups exhibits significant diversity of individual
source spectra (see the best-fitting spectral parameters for our
sources in \citealt{sazkha17a}). Nevertheless, the collective
spectra for these classes allow for some generalizing
description.

First, the collective spectrum of hard sources
resembles typical, hard spectra of 
ULXs in the so-called 'broadened-disc' and 'hard ultraluminous' states 
introduced by \citet{sutetal13}. To demonstrate this, we show in
Fig.~\ref{fig:avspec_spectypes_examples} the best-fitting (unabsorbed) model
(\citealt{sazetal14}, their table 2) for an X-ray spectrum of the
well-known ULX Ho IX~X-1 taken by the {\sl XMM-Netwon} observatory
(observation 0657801801), which consists of i) a hard ($\Gamma=1.28$) 
power-law component with a high-energy exponential cutoff at $E_{\rm
  cut}=6.4$~keV and ii) a weak additional, multicolour blackbody disc 
emission component with $kT_{\rm in}=0.3$~keV ({\sl cutoffpl}+{\sl diskbb}
in {\sc xspec}, \citealt{arnaud96}). We see that this spectrum nearly
matches our collective spectrum of hard sources.

Also shown in Fig.~\ref{fig:avspec_spectypes_examples} is the best-fitting
model (\citealt{sutetal13}, their table A1) for an {\sl XMM-Newton}
spectrum of NGC~5408~X-1, which according to the classification scheme
of these authors is a represenatative of the so-called 'soft
ultraluminous' ULX spectral class. In this case, the spectrum consists
of a fairly steep power-law component with $\Gamma=2.56$ and a
multicolour blackbody component with $kT_{\rm in}=0.194$~keV ({\sl
  powerlaw}+{\sl diskbb}). This spectrum is fairly similar, although
not an exact match, to our collective spectrum of soft sources.

As for the spectra of our supersoft sources, most of them can be 
described in terms of blackbody emission with $k\Tbb$ ranging from
$\sim 0.05$ to $\sim 0.25$~keV or multicolour disc emission with
$k\Tin$ ranging from $\sim 0.1$ to $\sim 0.3$~keV, although the
spectra of 5 supersoft sources are somewhat better described by a power law
with $\Gamma\sim 3.4$--3.8 (see \citealt{sazkha17a}). Hence, the
softer spectra in this category are similar to typical ULS spectra
\citep{diskon03,urqsor16} while the harder ones resemble the spectra
of 'normal' X-ray binaries in high/soft states associated with high
but subcritical accretion rates (see \citealt{donetal07} for a
review). Among our lowest luminosity ($10^{38}<\Lunabs\lesssim
  2\times 10^{38}$~erg~s$^{-1}$) supersoft sources there may also be
  present classical supersoft sources associated with accreting white
  dwarfs (e.g. \citealt{soretal16}), but such objects are not expected
  to provide a significant contribution to the collective spectrum of
  HMXBs, according to the HMXB XLF obtained in \citet{sazkha17a}.

Some or most of the harder spectra in the supersoft
  category may correspond
  to intermediate states between the supersoft ultraluminous state
  typical of ULSs and the soft ultraluminous state occuring in ULXs (see
  above). In fact, there is growing evidence that ultraluminous
  sources can make transitions between these states. One example of such
  behaviour is shown in Fig.~\ref{fig:avspec_spectypes_examples} based
  on the study by \citet{fenetal16}: NGC~247~ULS has
  been observed by {\sl Chandra} and {\sl XMM-Newton} to switch between
  i) a 'low', supersoft ultraluminous state, when its spectrum is
  dominated by soft thermal emission ($kT_{\rm in}=0.11$~keV) but
  exhibits an additional, weak power-law component ($\Gamma=2.1\pm
  0.9$, here we use the parameters for the {\sl diskbb}+{\sl powerlaw}
  model from table~3 of \citealt{fenetal16}), which dominates above
  $\sim 2$~keV, and ii) a 'high'  state, when the thermal component is 
  sowewhat harder ($kT_{\rm in}=0.15$~keV) and the power-law
  ($\Gamma=3.9\pm 0.4$) component provides a larger contribution to
  the X-ray luminosity\footnote{We have neglected an additional, weak 
    component representing thermal emission from an optically thin
    plasma in the best-fitting model of \citet{fenetal16} for
    NGC~247~ULS in its high state.}. The latter state appears to be
  intermediate between the supersoft and 
  soft ultraluminous states. A similar spectral transition has been
  observed in the well-known ULS in M101 \citep{sorkon16}. We see from
  Fig.~\ref{fig:avspec_spectypes_examples} that  our collective
  spectrum of supersoft sources may well be a superposition of
  spectra corresponding to different states of ULSs.

We conclude that the collective spectrum of HMXBs can be described in
terms of a superposition of different (known) spectral states of near-
and super-critically accreting X-ray binaries, which probably reflect
differences in the accretion rate and inclination of the accretion
disc with respect to the observer (see a further discussion in
\S\ref{s:discuss} below).

\subsection{Observed spectrum}
\label{s:avspec_obs}

\begin{figure}
\centering
\includegraphics[width=\columnwidth,viewport=30 200 560 710]{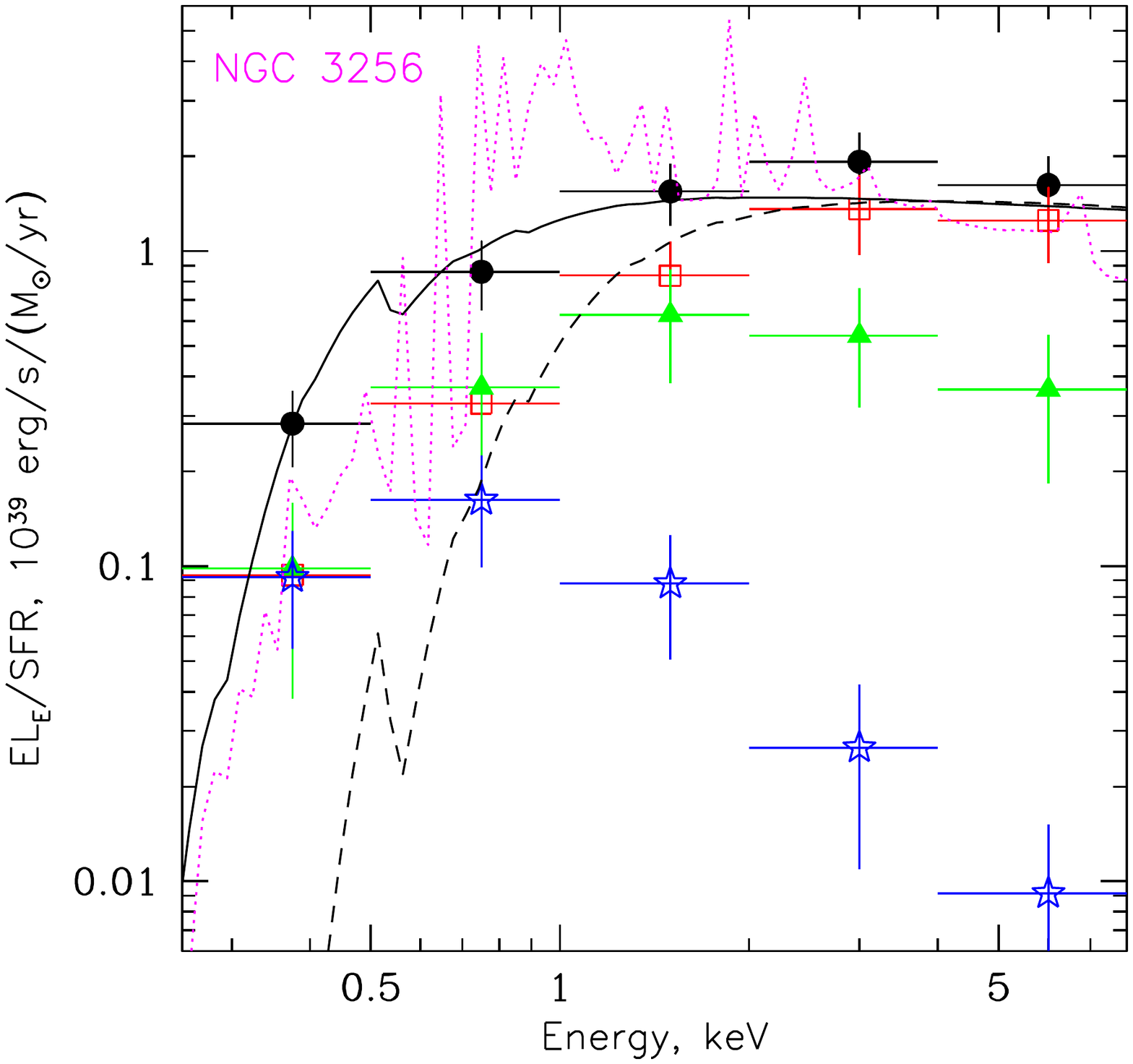} 
\caption{Observed collective spectrum of HMXBs
  ($10^{38}<\Lunabs<10^{40.5}$~erg~s$^{-1}$) (black
  circles) and the contributions of hard, soft and supersoft sources
  (squares, triangles and stars, respectively). The black solid line
  shows the best-fitting absorbed power-law model for the total
  spectrum, with a photon index fixed at $\Gamma=2.1$ (as inferred for
  the intrinsic HMXB spectrum) and $\NH=1.2\times
  10^{21}$~cm$^{-2}$. The magenta dotted line shows the observed
  SFR-normalized X-ray spectrum of the NGC~3256 starburst galaxy
  \citep{lehetal15}. The black dashed line shows our intrinsic 
    collective spectrum of HMXBs absorbed by $\NH=5\times
  10^{21}$~cm$^{-2}$ of cold gas, which is a crude estimate (based on
  the atomic and melecular gas content of this galaxy, see text) of
  the typical ISM column density towards the HMXBs in NGC~3256.
}
\label{fig:avspec_obs}
\end{figure}

Figure~\ref{fig:avspec_obs} shows the {\sl observed} collective
  spectrum of HMXBs, obtained using 
equation~(\ref{eq:avspec_obs}) by stacking the weighted spectra of all
200 sources in the sample
($10^{38}<\Lunabs<10^{40.5}$~erg~s$^{-1}$). As demonstrated in the
figure, the observed spectrum is dominated by hard sources, with the
contributions of soft and especially supersoft sources being
significantly suppressed compared to the intrinsic spectrum as a
result (mainly) of attenuation of their emission in the ISM of their host
galaxies (see \citealt{sazkha17a}). This effect is only noticeable
below 2~keV. The observed spectrum can be 
approximately described as the intrinsic spectrum [a power
  law with $\Gamma=2.1$, see eqs.~(\ref{eq:fit_l38_40.5},
  (\ref{eq:fit_l38_40})] absorbed in cold gas with column density
$\NH=(1.2\pm 0.2)\times 10^{21}$~cm$^{-2}$ (see
Fig.~\ref{fig:avspec_obs}). This value is very close to the median 
absorption column of $1.1\times 10^{21}$~cm$^{-2}$ for our sample of
sources, inferred from their X-ray spectra \citep{sazkha17a}. 
 
It is interesting to compare the observed collective spectrum of
  HMXBs constructed here with observed galaxy-wide
X-ray spectra of actively star forming galaxies. Several such
spectra, measured with {\sl Chandra} and {\sl NuSTAR}, have been
presented by \citet{lehetal15}, of which the most interesting is that
of the starburst galaxy NGC~3256. This galaxy has a very high total
SFR of $\sim 36$~$M_\odot$~yr$^{-1}$, which is similar to the combined
SFR of all 27 galaxies making up our sample ($\sim
32$~$M_\odot$~yr$^{-1}$). Therefore, NGC~3256 should be as
representative of the local luminous HMXB population as our sample of
galaxies.

As shown in Fig.~\ref{fig:avspec_obs}, the SFR-normalized spectrum of
NGC~3256 nearly matches our observed collective spectrum of HMXBs at
energies above $\sim 3$~keV, confirming that the total emission of actively
star forming galaxies at these energies is dominated by ULXs. It is
more difficult to compare the collective spectrum of HMXBs and
the NGC~3256 spectrum below 3~keV. First, as discussed by
\citet{lehetal15} and evident from the strong X-ray line emission
observed from NGC~3256, thermal emission from hot interstellar gas
provides a strong 
contribution to the soft X-ray flux from this galaxy. Secondly, soft
X-ray emission from point sources in NGC~3256 can be significantly
suppressed by absorption in the cold component of its ISM, in fact
much stronger than for typical galaxies in our sample, which are
characterized by much lower SFRs (at most $\sim
3$~$M_\odot$~yr$^{-1}$) and smaller amounts of cold gas.

We can roughly estimate the expected absorption column density for the
X-ray sources in NGC~3256 using measurements of its total gas content
and assuming that the gas is uniformly distributed over a disc of some
characteristic radius $R$. For the atomic gas, the total mass is
estimated as $M_{\rm HI}\sim6\times 10^{9}~M_{\odot}$, with $R_{\rm
  HI}\sim 30$~kpc (e.g. \citealt{casetal04}), while for the molecular
gas $M_{\rm H_2}\sim5\times 10^{9}~M_{\odot}$ with $R_{\rm H_2}\sim 6.6$~kpc
\citep{uedetal14}. This yields integrated column densities
(perpendicular to the plane of the galaxy) of $\NH\sim 3 \times
10^{20}$ and $\sim 5\times10^{21}$~cm$^{-2}$ for the atomic and
molecular gas, respectively. Taking into account that NGC 3256 is
inclined at $\sim 48^\circ$ (according to
HyperLeda\footnote{http://leda.univ-lyon1.fr/}) to our line of sight
and the Galactic absorption of $\sim 7\times10^{20}$~cm$^{-2}$ in its
direction \citep{kaletal05}, we infer that X-ray sources in this
galaxy should typically be screened from us by $\NH\sim 5\times
10^{21} $~cm$^{-2}$ of cold gas. Subjecting our intrinsic
($\Gamma=2.1$) collective spectrum of HMXBs to this amount of absorption
results in a spectrum shown in Fig.~\ref{fig:avspec_obs}. We see that
most of the soft X-ray emission produced by the HMXBs in NGC~3256 can
be obscured by the ISM. In reality, HMXBs are usually concentrated to
regions of active star formation and enhanced gas column density, so
the ensemble-averaged $\NH$ for the HMXB population of NGC~3256 can be
even higher than in our estimate.

Comparison of Fig.~\ref{fig:avspec_lumranges} and
Fig.~\ref{fig:avspec_obs} suggests that the (mostly obscured)
population of luminous soft HMXBs in NGC~3256 probably produces a
similar amount of soft X-rays as its hot interstellar gas. This
  is in agreement with the conclusion reached by 
  \citet{minetal12a,minetal12b}, who compared the integrated
  contributions of point X-ray sources and hot ISM to the galaxy-wide
  X-ray luminosity for two dosens of nearby galaxies and found both
  contributions to be similar and proportional to the SFR,
  albeit within a large uncertainty associated with cold-gas
  absorption of soft X-rays emitted by hot gas. According to the
  linear relation found by \citet{minetal12b}, the 
  intrinsic X-ray (0.3--10~keV) luminosity of the ISM of a galaxy with
  a given SFR is expected to be $\sim 7\times 10^{39}\,({\rm
    SFR}/M_\odot~\rm{yr}^{-1})$~erg~s$^{-1}$. Taking into account the
  substantial intrinsic absorption in NGC~3256 (see the discussion
  above) it appears from Fig.~\ref{fig:avspec_obs} that NGC~3256
  is consistent with the \citet{minetal12b} correlation.

The apparent approximate parity between the total X-ray outputs of the
luminous HMXB population and hot ISM in star forming galaxies is
interesting and should be further studied in future work.  

\section{Discussion}
\label{s:discuss}

Although the exact nature of the sources comprising our sample is
unknown, we have demonstrated that the intrinsic collective
  spectrum of HMXBs can be described in terms of a superposition of
various known spectral states of luminous X-ray binaries: the
broadened-disc, hard ultraluminous and soft ultraluminous states known
for ULXs \citep{sutetal13}, the very soft ($\Tbb\sim 0.1$~keV)
blackbody-like state typical of ULSs \citep{urqsor16} and the high/soft
states of 'normal' X-ray binaries \citep{donetal07}. All these states
may be different manifestations of near- or super-critical accretion
of matter from a massive stellar companion onto a stellar-mass black
hole (or a neutron star in some systems), reflecting differences in
the accretion rate and/or in the orientation of the (thick) accretion
disc and its wind with respect to the observer
(e.g. \citealt{midetal15,fenetal16,guetal16,urqsor16}). The basic idea
discussed in these recent papers is that when the disc is observed
nearly face-on, (relatively) hard X-ray radiation from the central
funnel is directly visible. However, the central emission region can be
obscured by the wind from an observer viewing the disc at larger
inclination, so that only reprocessed softer emission will be visible. 

\citet{kawetal12} have performed a detailed modelling of X-ray
spectra generated by supercritical accretion onto a stellar-mass black
hole, combining the results of hydrodynamical simulations of a
thick accretion flow with Monte-Carlo radiative transfer in this
flow. The findings of this work are in good agreement with the general
picture adopted in the aforementioned studies, namely the appearance
of a supercritical accretor should strongly depend on the viewing
angle: the source will appear more luminous and harder if observed
face-on and weaker and softer if observed at a large angle. According
to the angular dependence of the observed X-ray luminosity (for a given
accretion rate) obtained by \citet{kawetal12} (see their fig.~3),
objects viewed at intermediate angles of $i\sim 10$--$40^\circ$ should
dominate in the collective X-ray emission of the local population of
supercritical accretors (assuming, of course, that they are randomly
oriented). Therefore, the angle-integrated spectrum of such sources 
should be somewhat softer than the spectra of face-on ($i=0$) objects,
namely it is expected to have an effective photon index of $\Gamma\sim
2$ according to fig.~4 in \citet{kawetal12}. This value is close to
the $\Gamma=2.1\pm 0.1$ slope of our collective
spectrum of luminous HMXBs, demonstrating that this
spectrum (as well as its composition in terms of sources of different
luminosities and spectral types) places interesting observational
constraints on supercritical accretion models.

The present study suggests that the average spectral hardness of
  luminous HMXBs does not strongly depend on their luminosity (compare
  the collective spectra for three different luminosity bins in
  Fig.~\ref{fig:avspec_lumranges}), which seems to contradict
  the general picture outlined above, according to which spectral
  hardness should positively correlate with the observed
  luminosity of ULXs. Part of the explanation why the collective
  spectrum of our least luminous
  ($\Lunabs=10^{38}$--$10^{39}$~erg~s$^{-1}$) sources is as hard as
  the spectra of more luminous objects may be that this low-luminosity 
  bin probably includes, apart from (soft) supercritical accretors,
  sub- and near-critically accreting black holes and neutron stars with 
  relatively hard spectra. Another reason may be that the spectrum
  for this luminosity bin is significantly affected by LMXB
  contamination, which we have roughly taken into account [via
  eq.~(\ref{eq:lmxb})] for the normalization but not for the shape
  of the spectrum. 

According to the collective spectrum of HMXBs, soft X-ray
emission (0.25--2~keV) dominates the total radiative output of HMXBs in the 
local Universe. This fact has been frequently overlooked in previous
studies, because much of this soft X-ray emission is absorbed in the
ISM and does not reach the Earth's absorber. The lower energy boundary
of 0.25~keV in our analysis is mainly set by the sensitivity of the {\sl 
  Chandra} X-ray telescope. What if the collective spectrum of
  HMXBs continues with nearly the same slope ($\Gamma=2.1$) to yet
lower energies? This would mean that the luminous 
HMXB population produces, apart from X-rays, a comparable or even higher
luminosity, $\gtrsim 5\times
10^{39}$~erg~s$^{-1}(M_\odot$~yr$^{-1})^{-1}$, at UV and lower
frequencies. This hidden radiation, if real, may be associated with
'misaligned ULXs', i.e. supercritically accreting massive binaries
viewed at yet higher inclinations and/or having yet higher
accretion rates than ULXs and ULSs, and the famous Galactic
microquasar SS~433 may be one of such systems
(e.g. \citealt{fabrika04,pouetal07,khasaz16}). Indeed, the observed
(albeit only at $E\gtrsim 2$~keV, because of strong line-of-signt
absorption in the soft band) X-ray luminosity of SS~433 is only 
$\sim 10^{36}$~erg~s$^{-1}$ (and it is associated with its baryonic jets
rather than directly with the central source), while its UV luminosity
is estimated as $\sim 10^{40}$--$10^{41}$~erg~s$^{-1}$
\citep{cheetal82,doletal97}, and this radiation is probably associated
with the photosphere of the disc wind \citep{fabrika04}.

\section{Conclusion}
\label{s:summary}

Using a sample of 200 luminous ($\Lunabs>10^{38}$~erg~s$^{-1}$) HMXB
candidates detected by {\sl Chandra} in 27 nearby galaxies, we have
constructed the collective X-ray spectrum of HMXBs in the local
  Universe per unit star formation rate, corrected 
for observational biases associated with intrinsic diversity of source
spectra and X-ray absorption in the ISM (of the host galaxies and
the Milky Way). This spectrum can be described by a power law with a
photon index $\Gamma=2.1\pm 0.1$ [eqs.~(\ref{eq:fit_l38_40.5}) and
  (\ref{eq:fit_l38_40})] and is dominated by ultraluminous sources
($\Lunabs>10^{39}$~erg~s$^{-1}$), with comparable contributions from
hard, soft and supersoft sources [as defined in
  eq.~(\ref{eq:types})]. Hard sources, whose spectra resemble those of
'classical' ULXs, dominate at energies above a few keV, while the bulk
of the soft X-ray emission (below 2~keV) is provided by soft and
supersoft sources.

If our favoured interpretation that the derived spectrum mainly
represents population- and angle-integrated emission from
supercritically accreting HMXBs is correct, then its nearly flat
shape (in $\nu F_\nu$ units) provides an interesting constraint on
theoretical models of supercritical accretion. 

The strong soft X-ray emission revealed by the intrinsic collective
  spectrum of HMXBs could play an important role in the early
Universe, since the ISM in the first galaxies was probably more
transparent to X-rays than in present-day galaxies, in particular due
to the lower metallicity of the former. As a result, soft and
supersoft luminous HMXBs might have been the key contributors to the
X-ray heating of the Universe prior to its reionization, as we have
discussed recently \citep{sazkha17b}. The collective spectrum of
  HMXBs obtained here can thus be used as a reference spectrum for
detailed simulations of cosmic X-ray preheating.

\section*{Acknowledgments}

The authors thank the referee for useful suggestions. 
  

\bsp	
\label{lastpage}
\end{document}